# Thermophoretic Levitation of Solid Particles at Atmospheric Pressure

Pritam Kumar Roy,[1] Irina Legchenkova,[1] Leonid A. Dombrovsky[1,2,3], Vladimir Yu. Levashov[4], Alexei P. Kryukov[5], Bernard P. Binks[6], Nir Shvalb[7], Shraga Shoval[8], Viktor Valtsifer[9], Edward Bormashenko[1*]

[1]Department of Chemical Engineering, Faculty of Engineering, Ariel University, Ariel 407000, Israel
[2]Heat Transfer Department, Joint Institute for High Temperatures, 17A Krasnokazarmennaya St., Moscow 111116, Russia
[3]X-BIO Institute, University of Tyumen, 6 Volodarskogo St., Tyumen 625003, Russia
[4]Institute of Mechanics of Moscow State University, 1 Michurinskiy Prosp., Moscow 119192, Russia
[5]Moscow Power Engineering Institute, 14 Krasnokazarmennaya St., Moscow 111250, Russia
[6]Department of Chemistry, University of Hull, Hull HU6 7RX, UK
[7]Department of Mechanical Engineering & Mechatronics, Faculty of Engineering, Ariel University, P.O.B. 3, Ariel 407000, Israel
[8]Department of Industrial Engineering and Management, Faculty of Engineering, Ariel University, P.O.B. 3, Ariel 407000, Israel
[9]Institute of Technical Chemistry, Academician Korolev St., 3, Perm 614013, Russia.

[*]Corresponding Author: Edward Bormashenko edward@ariel.ac.il

## Abstract

Separation and transportation of small particles are important processes in various applications such as the food and pharmaceutical industries. Although mechanical, chemical or electrical methods can provide possible solutions, operational or environmental constraints may require alternative methods. Spreading and levitation of clusters (aggregates) of fumed silica nanoparticles placed under atmospheric pressure on a hot plate is reported. In a closed chamber, the particles started to spread horizontally at the threshold temperature of $T_c^* = 403 \pm 1$K. The powder spreading in the chamber continued until the temperature-dependent saturation value $r_{sat}(T)$, which grew linearly with the temperature. Open space experiments clearly demonstrated levitation of the powder clouds. The onset of levitation in the open space corresponded to the minimal threshold temperature of $T_o^* = 373 \pm 1$K. Qualitative physical analysis of the observed phenomena is suggested. The effect of levitation is explained by the lifting thermo-phoretic force emerging in the Knudsen layer of air on the heater surface. The levitation of the powder under atmospheric pressure becomes possible due to the combination of low adhesion of the fluorinated fumed silica clusters built of nanoparticles to the substrate, relatively low density of the particles and clusters, and their high specific surface area. Ordering of the aggregates of nanoparticles within the levitating powder cloud was quantified with Voronoi diagrams.

Several physical effects allow stable levitation, namely, the process in which an object is held aloft without mechanical support.[1] The noun levitation emerges from the Latin word "levitas" which means "lightness". Although magicians have perfected sleight of hand to demonstrate free gliding physical bodies, physicists have studied a diversity of phenomena enabling true levitation of condensed matter (liquid or solid)[1]: aerodynamic and acoustic levitation[2-5], optical levitation[6-7], electromagnetic levitation[8-9] and superconducting levitation[10]. The phenomenon which is adjacent to levitation is the so-called Leidenfrost effect.[11-15] The Leidenfrost effect is a phenomenon in which a liquid (droplet), in close contact with a solid body that is significantly hotter than the liquid's boiling point, produces an insulating vapor layer which keeps that liquid from boiling rapidly.[11] Leidenfrost drops are very mobile.[11-13,16] Not only liquids but also volatile solids demonstrate Leidenfrost-like levitation and self-propulsion.[17] Levitation of dust aggregates composed of micrometer-sized grains from 100 µm to 1 cm in size moving over a hot surface was demonstrated in ref. 18. Depending on the dust sample, the aggregates start to levitate at a temperature of 400 K. Levitation of dust aggregates is restricted to a pressure range between 1 and 40 mBar. The levitation is related to the Knudsen thermo-phoretic lifting force.[18-23] Theoretical analysis of the Knudsen layer assisted levitation is performed in ref. 23. We demonstrate here that strongly hydrophobic powders may spread and levitate above a hot surface when driven by the thermo-phoretic force under atmospheric pressure and relatively low temperatures of the surface ($T \geq 373 \pm 1K$).

*Closed space experiments: spreading of fluorinated fumed silica powder*

We examined the behavior of a diversity of nanopowders (see the **Methods Section**) heated from below in the closed experimental cell shown in **Extended Data** Fig. 1a at atmospheric pressure as well as in the open space of the laboratory as depicted in **Extended Data** Fig. 1b. We heated from below a diversity of organic and nonorganic nanopowders, including fluorinated fumed silica, PTFE, lycopodium and silica powder in a closed space. Fumed silica nanopowder was synthesized as described in the **Methods Section** and ref. 18. Fluorinated fumed silica demonstrated a surprising physical behavior when heated. At the threshold temperature of $T_c^* = 403 \pm 1$K (subscript "$c$" denotes closed), fluorinated fumed silica powder started to spread along the heated surface, as shown in Figs. 1 (a-b) and Movie S1. The spreading powder formed an approximately circular "cloud" built of the agglomerates (clusters) of the fluorinated fumed silica particles. The average size of the nonspherical clusters (aggregates) of nanoparticles was $\bar{d} \cong 0.2 \pm 0.05$ mm ($\bar{d} = \frac{\sum_{i=1}^{N} d_i}{N}$, where $d_i$ is the dimension of the $i$-th cluster).

It should be emphasized that other powders remained at rest when heated. It is difficult to answer the following question: is the spreading of the powder accompanied by its levitation? The levitation height in the closed chamber was very small (less than one micrometer) and it turned out to be indistinguishable with our experimental facilities.

Quasi-isotropic spreading of the fluorinated fumed silica powder was quantified, as shown in Figs. 1 and 2. Initially, the circular spot of the fluorinated fumed silica powder increased its initial radius *r* with time upon heating the substrate, as shown in Fig. 2a. Isothermal spreading of the spot at the fixed temperature *T* in turn continued until the temperature-dependent saturation value $r_{sat}(T)$, presented in Fig. 2b. The dimensionless dependencies $\hat{r}(\hat{T})$ and $\hat{r}_{sat}(\hat{T})$ are well-fitted linearly with Eqs. (1a-b):

$$\hat{r}(\hat{T}) = \alpha + (1 - \alpha)\hat{T} \qquad (1a)$$

$$\hat{r}_{sat}(\hat{T}) = \gamma + (1 - \gamma)\hat{T} \qquad (1b)$$

where $\hat{r}(\hat{T}) = \frac{r(\hat{T})}{r_0}$; $\hat{r}_{sat} = \frac{r_{sat}(\hat{T})}{r_{sat0}}$ and $\hat{T} = \frac{T}{T_c^*}$ are the dimensionless radius of the spot and the temperature correspondingly, $r_0 = r(T < T_c^*)$; $r_{sat0} = r_{sat}(T = T_c^*)$; $\alpha$ and $\gamma$ the dimensionless fitting constants. Phenomenological explanation of the physical origin of Eqs. 1a-b is suggested below.

Now we address the following question: how does the saturation radius of horizontal spreading given by Eq. (1a) emerge? We use the Jeans-like model of spreading of a cloud, rooted in celestial mechanics and take into account electrostatic interactions between agglomerates.[23-24] Consider a homogenous cylindrical cloud of clusters with an initial radius $r$, shown in Extended Data Fig. 3. Assume that the clusters within the cloud interact *via* a phenomenological attractive potential (say electrostatic-like potential[25]): $U = k\frac{Q}{L}\ln\frac{r}{\hat{r}_0}$, where $Q$ and $\hat{r}_0$ are phenomenological parameters, the dimensions of $Q$ are those of the electrical charge, and $\hat{r}_0$ is the cut-off length which is reasonably equaled to $\bar{d}$, $k$ is the units-dependent constant and $L$ is the thickness of the cloud. This kind of potential is typical for problems characterized by the cylindrical symmetry (see **Extended Data** Fig. 3).[26] At the saturation stage, the value of the thickness of the cloud was established experimentally as $L \cong 0.3 \pm 0.05$ mm; in other words, it was close to the characteristic size of a single cluster. In order to compress this cylinder to a radius $r - dr$, work must be done against the "gas pressure". During the compression, energy $dU$ is released:

$$dU = k\frac{Q}{L}\frac{1}{r}dr \qquad (2)$$

When this energy equals the amount of work $dW$ to be done on the gas of clusters, the critical (saturation) radius $r_{sat}$ is attained. The work $dW$ is estimated as:

$$dW = n(T)T2\pi rLdr, \qquad (3)$$

where $n(T)$ is the temperature-dependent concentration of clusters in the cloud. Equating expressions (2) and (3) and assuming $n(T)L\pi r^2 = N = const$ (where $N$ is the total number of clusters), we obtain the saturation radius:

$$r_{sat}(T) = N\sqrt{\frac{2}{\pi k n(T) Q}}\sqrt{T} \tag{4}$$

One can easily extract the concentration radial distribution of the cloud, shown in Fig. 1d:

$$\frac{n(r)}{N} = \frac{r_{sat}}{2\pi r^3 L} \tag{5}$$

If the model of the "ideal gas" is adopted for the cloud, the concentration of clusters is given by:

$$n(T) = \frac{P_{air}}{k_B T}, \tag{6}$$

where $P_{air} = const$ is the atmospheric pressure and $k_B$ is the Boltzmann constant. Substitution of Eq. (6) into Eq. (4) yields the experimentally observed scaling law, namely $r_{sat} \sim T$, shown in Fig. 2b.

### *Open space experiments: spreading of fluorinated fumed silica powder*

The open space experiments, illustrated in Fig. 1b, generated even more surprising, fascinating experimental findings. Spreading of the fluorinated fumed silica powder was accompanied with its levitation, as shown in **Extended Data** Figs. 1b, 3, 4. The altitude (levitation height) of the cloud built of the clusters (the gap between the cluster and the substrate), denoted $h$ and shown in **Extended Data** Figs. 1b, 3 and Movie S2, was $h = 15 - 18 \pm 1$ μm which changed slightly during the levitation (for details of the experimental procedure which enabled the measurement of $h$, see the **Methods Section**). It should be mentioned that the value of $h$ was slightly temperature-dependent. Motion of the entire cloud is depicted in Fig. 4, **Extended Data** Fig. 4, Movie S3 and S4. The temperature corresponding to the onset of levitation in the open space was established as $T_o^* = 373 \pm 1$K; thus, the interrelation $T_o^* < T_c^*$ takes place.

The observed difference between the threshold temperatures of levitation obtained within the closed chamber and in the open space indicated that a part of the heat supplied to the particle cloud is spent to the chaotic horizontal motion of the cloud in the open space. The heat flux to the cloud in the Knudsen layer increases with the hot plate temperature. At the same time, according to the generalized le Chatelier's principle, the relatively stable cloud tends to maintain its temperature. This could be partially achieved by heat transfer to the air above the cloud, but it is insufficient because of the low thermal conductivity of air. Fortunately, there remains the possibility of converting a part of the heat into a kinetic energy of the cloud motion along the boundary of the Knudsen layer. This motion, with an average velocity increasing with temperature, was observed in the experiments.

Inside the closed chamber, the motion of viscous air is partially suppressed. This prevents the motion of the cloud as well. As a result, the thermal energy cannot be dissipated by the motion of the cloud. Most likely, the difference in the threshold temperatures of levitation between the closed and open experiment settings is due to this physical phenomenon.

The average size of the non-spherical clusters (aggregates) of nanoparticles within a cloud was $\bar{d} \cong 0.2 \pm 0.05 \ mm$ ( $\bar{d} = \frac{\sum_{i=1}^{N} d_i}{N}$, where $d_i$ is the dimension of the $i$-th cluster). The double averaged distance between levitating solid clusters was $\bar{\bar{s}} = 0.7 \pm 0.05$ mm (the mathematical procedure is described in the **Methods Section**); thus, the inter-relation $\frac{\bar{d}}{\bar{\bar{s}}} \leq 1$ takes place. It is also noteworthy that the inter-relations $\frac{h}{\bar{d}} \ll 1; \frac{h}{\bar{\bar{s}}} \ll 1$ took place under the levitation. In other words, the levitation height was much smaller than both the characteristic size of the clusters and the averaged distance between the levitating clusters. Motion of the entire cloud in the open space was random and depended on the weak air flows, inevitable in the laboratory.

We relate the origin of levitation to the thermo-phoretic force acting on the cluster occurring in the Knudsen layer. At the beginning of the heating process, the distance between the lower boundary of the cluster and the hot plate is comparable with the mean free path of air molecules. It means that the lifting force occurs under conditions of the free molecular (Knudsen) flow supporting the cluster. The physical source of the lifting force is the momentum transferred to the cluster by relatively fast molecules of air moving from the hot plate almost without colliding with other molecules. The main parameter of the free molecular flow is the Knudsen number:

$$\text{Kn} = \lambda/h \tag{7}$$

where $\lambda$ is the mean free path of molecules. Note that $\lambda \approx 0.07$ µm for air molecules at normal atmospheric conditions. This value increases up to $\lambda \approx 0.1$ µm when the air temperature is equal to $T_o^* = 373 \pm 1$K.

The lifting force acting on the cluster decreases with increasing the height of levitation because of a transition to the regime of a continuous medium at Kn≪1, when fast molecules of air do not reach the cluster due to numerous collisions between the molecules, whereas the viscous forces still do not allow the upward gas flow to develop and the lifting force disappears. The latter means that the cluster cannot completely leave the Knudsen layer. It should be noted that the size of the nanoparticle aggregates is much greater than the free path of air molecules, and the upper parts of the aggregates are above the Knudsen layer. Nevertheless, one can say that the Knudsen layer is indirectly observed in the experiment.

The accurate solution of the problem under consideration should be based on the kinetic Boltzmann equation, which can be solved using present-day numerical methods.[27-30] One of the low-order differential approximations, like the four-moment method of the early paper,[31] can be also employed to obtain an analytical solution which is convenient for the analysis of the integral parameters such as the gas pressure and heat flux. In further qualitative consideration, we will restrict ourselves to the simplest relations for the heat flux, $q$, and the lifting pressure, $\Delta p$, for the levitating plate considered instead of the particle cloud:

$$q = 4 p_{air} \sqrt{\frac{R_{air}}{2\pi}} \left( \sqrt{T} - \sqrt{T_{cl}} \right) \tag{8}$$

$$\Delta p = p_{air} \left( \sqrt{T/T_{cl}} - 1 \right), \tag{9}$$

where $T_{cl}$ is the temperature of the cloud and $R_{air} = 287$ J/(kg K) is the gas constant of air. It is assumed here that the temperature of the small aggregates is very close to the local temperature of the ambient air. A convergence of a more general solution based on the method reported in ref. 31 to the limiting solution (9) at a large Knudsen number is illustrated in Fig. 5 of the **Extended Data**. One can see that Eq. (9) can be used to calculate $\Delta p$ in the case of $T_{cl} \approx T$.

At the initial stage of levitation, the Knudsen lifting force should overcome the adhesion between the cluster and the hot plate. The modulus of the van der Waals (adhesion) force can be calculated as follows:

$$F_{adh} \cong \frac{AR}{6D^2}, \tag{10}$$

where $A \approx 10^{-19}$J is the Hamaker constant,[32] $R$ is the characteristic dimension of the cluster $\left(R \cong \frac{\bar{d}}{2}\right)$, and $D$ is the size of the gap separating the particles from the hot substrate. Equating the Knudsen lifting force to the adhesion force yields:

$$p_{air}\left(\sqrt{\frac{T}{T_{cl}}} - 1\right)\frac{\pi \bar{d}^2}{4} = \frac{A\bar{d}}{12D^2} \tag{11}$$

Eq. (11) is easily re-shaped as follows:

$$\frac{T}{T_{cl}} = (1 + \xi)^2, \tag{12}$$

where $\xi = \frac{A}{3\pi \bar{d} D^2 p_{air}}$ is the dimensionless parameter defining the possibility of detachment of a cluster from the hot plate. Assuming $A \cong 10^{-19}$J, $D \cong 0.2$ nm, $\bar{d} \cong 0.2$ mm, $p_{air} \cong 10^5$ Pa yields $\xi \cong 10^{-2}$, Eq. (12) can be written as follows:

$$\frac{T}{T_{cl}} - 1 \cong 2\xi \cong 2.0 \times 10^{-2} \tag{13}$$

Thus, even the relatively small temperature difference between what and what? provides detachment of the clusters from the hot plate. At the next (levitation) stage, the Knudsen lifting force should withstand gravity, and the pressure difference supporting a single aggregate of nanoparticles can be estimated as:

$$\Delta p \cong \frac{2}{3}\rho_{eff}\bar{d}g \tag{14}$$

where $\rho_{eff}$ is an effective density of the agglomerates; assuming as a packaging factor of agglomerates $\Omega \cong 0.45 - 0.6$, we estimate $\rho_{eff} = \Omega\rho_s$ where $\rho_s = 2200$ kg/m³ is the density of the fumed silica particles; thus $\rho_{eff} \cong (1.3 - 1.0) \times 10^3$ kg/m³ supplies a reasonable estimation for the effective density of agglomerates. Note that the value of $\Delta p$ in the Knudsen layer does not depend on the height of the cluster levitation. Substituting $\rho_{eff}$ we obtain $\Delta p = 2.0 - 2.7$ N/m². As one can expect, this value is very small and $\Delta p \ll p_{air}$. According to Eq. (11) (see also Fig. 5 of the **Extended Data**), this means that the cloud temperature is only slightly less than the temperature of the hot plate.

A levitating particle should surmount two forces, namely: the van der Waals force and gravity. Let us estimate a very important length scale $H$ at which these forces become comparable namely:

$$\frac{AR}{6H^2} \cong \frac{4}{3}\rho\pi R^3 g \tag{15}$$

For the sake of a very rough estimation, we obtain:

$$H \cong \frac{1}{5R}\sqrt{\frac{A}{\rho g}} \tag{16}$$

Assuming $R = 25 \times 10^{-9}$m, $A = 10^{-19}$J yields $H \cong 18.0$ μm. It is seen that the height $H$ arising from Eq. (16) coincides more than satisfactorily with the experimentally established levitation height. However, this is true only for levitating isolated nanoparticles; obviously at the height of $H \cong 18.0\ \mu m$ gravity dominates over the van der Waals forces acting on levitating clusters.

It should be noted that both spreading and levitation of the fumed fluorosilica clusters were observed at atmospheric pressure on various hot substrates including glass surfaces, aluminum foil, iron surfaces and ceramic surfaces as shown in Movie S5, S6, S7.

Let us pose the following fundamental question: are the clusters within a cloud ordered? The answer to this question is supplied by the time-evolution of the Voronoi entropy of the cloud, denoted $S_{vor}$ (see Fig. 6 of the **Extended Data**).[33-34] The Voronoi entropy of the cloud was calculated as explained in the **Methods Section**. The value of the Voronoi entropy was established as $S_{vor} = 1.4 \pm 0.2$, which is slightly lower than the value corresponding to the random 2D pattern, $S_{vor} = 1.71 \pm 0.2$. Thus, the observed cloud may be defined as slightly ordered, the physical reasoning of the ordering calls for additional physical insights.

We tested a number of nanopowders, but only the fluorinated fumed silica powder demonstrated the effects of spreading and levitation. The reasonable question is: what is special with this kind of powder? We suggest that the observed effects of spreading and levitation are due to the combination of the low value of the Hamaker constant,[32] relatively low density and unusually high specific surface area, enabling the Knudsen-force-inspired detachment of the powder from the hot plate followed by its displacement by air flows appearing as levitation.

We demonstrated experimentally the spreading and levitation of aggregates of fluorinated fumed silica nanoparticles placed on the hot substrate. It should be emphasized that both spreading and levitation were observed at atmospheric pressure. Various findings were observed in open and closed space experiments. The isotropic spreading of the powder cloud was observed in the closed space. The onset of spreading of the powder in the closed space experiments corresponded to the temperature $T_c^* = 403.0 \pm 1 \text{K}$. The radius of the spreading cloud grew linearly with the temperature of the hot substrate and came to a saturation. A phenomenological physical model, based on the "ideal gas model" of the cloud predicting the saturation radius, is suggested.

Levitation of the clouds under investigation was observed in open space experiments. The levitation is explained by the lifting thermo-phoretic force emerging in the Knudsen layer supporting the powder. The levitation height was established as $h = 18 \pm 1$ μm and it corresponds to the small Knudsen number. Levitation becomes possible when the Knudsen thermo-phoretic force is larger than the van der Waals adhesion between the substrate and the powder. Ordering of levitating clusters built from nanoparticles was quantified with Voronoi diagrams. The observed effect can be used to separate some powders, exploiting the levitation of certain kinds of powders. It may also be used for the non-contact transportation of such powders. This method can be beneficial in applications where mechanical/chemical/electrical methods cannot be used, such as in the food or pharmaceutical industries.

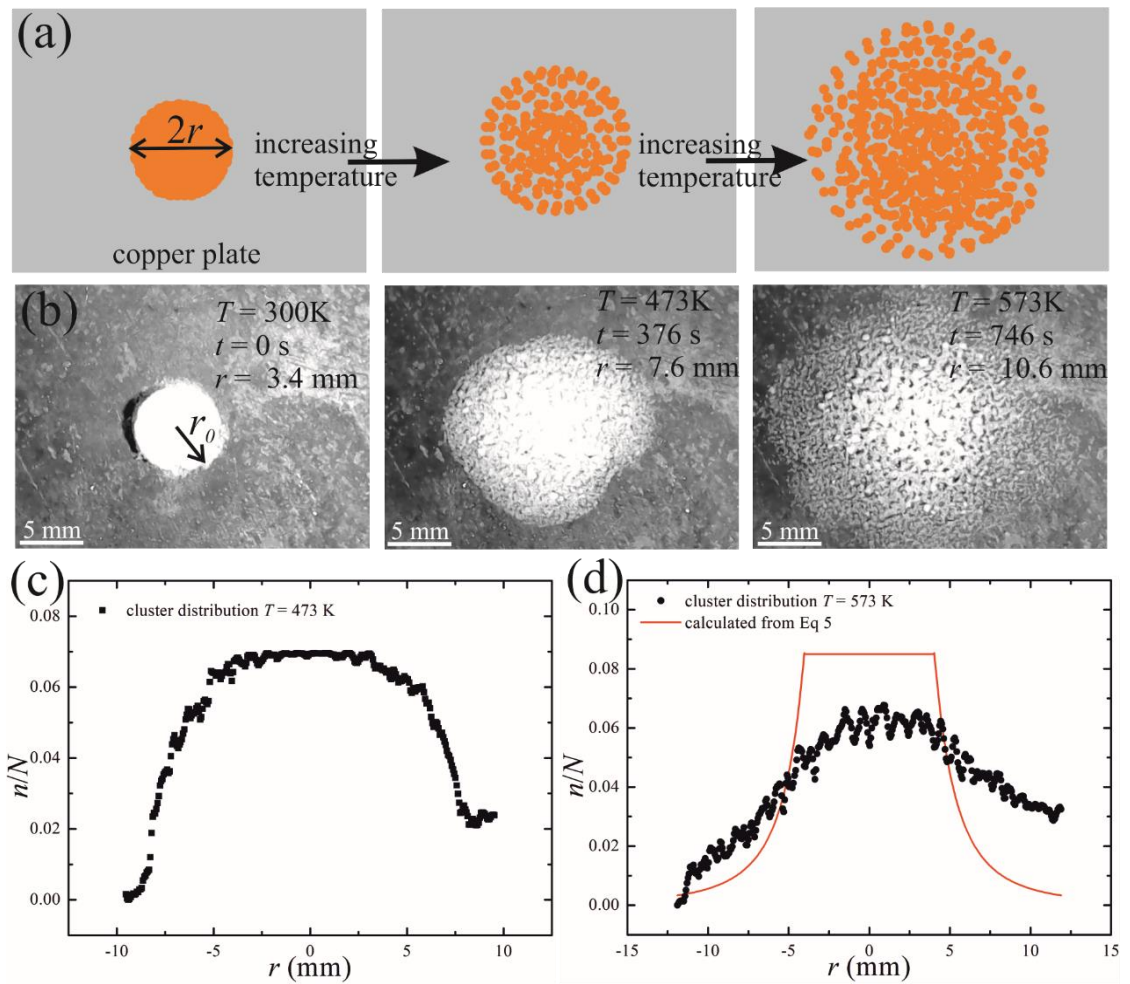

Fig. 1 | **Spreading of fluorinated fumed silica nano-powder in a closed chamber is illustrated. a,** Schematic. **b,** Sequence of images. Initial radius at temperature $T = 300$ K; $r = 3.4 \pm 0.5$ mm and the radius of the cloud at temperature $T = 573$ K; $r = 10.6 \pm 0.5$ mm. **c.** Experimentally established at $T = 473$ K radial concentration distribution of the agglomerates in the cloud $\frac{n(r)}{N}$ is depicted; **d.** Comparison of the experimentally established at $T = 573$ K radial concentration distribution of the cloud $\frac{n(r)}{N}$ with that calculated with Eq. 5 is shown.

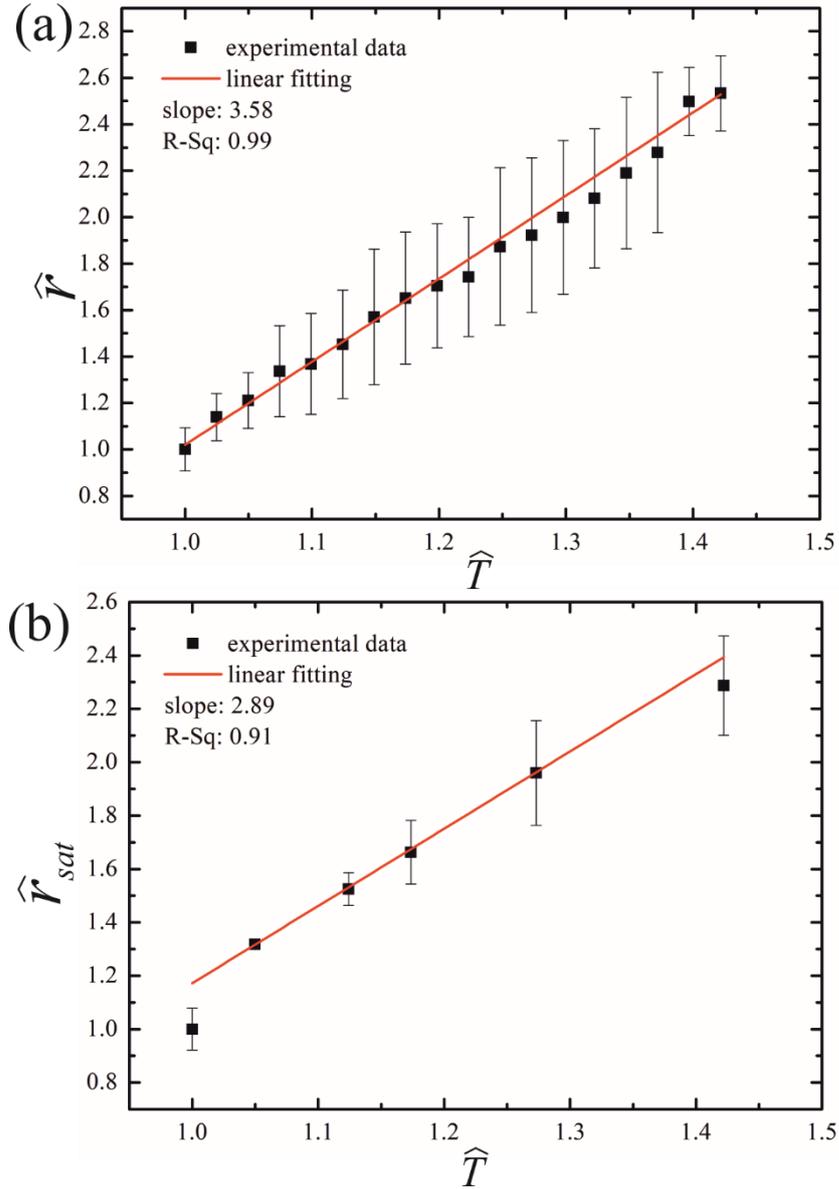

**Fig. 2 | Dimensionless spreading and saturation radius of the fumed silica cloud as a function of dimensionless temperature. a,** Dimensionless spreading radius ($\hat{r}(\hat{T}) = \frac{r(\hat{T})}{r_0}$) of the fumed silica cloud as a function of dimensionless temperature ($\hat{T} = \frac{T}{T^*}$) is depicted; $r_0 = r(T < T^*)$; $T^* = 403 \pm 1$ K; $r_0 = 3.6 \pm 0.5$ mm. Solid red line represents the linear fitting: $\hat{r}(\hat{T}) = \alpha + (1-\alpha)\hat{T}$; $\alpha = 3.58$ **b,** Dimensionless saturation radius ($\hat{r}_{sat} = \frac{r_{sat}(\hat{T})}{r_0}$) of the cloud as a function of dimensionless temperature ($\hat{T} = \frac{T}{T^*}$) is depicted; $r_{sat0} = r_{sat}(T = T^*) = 4.4 \pm 0.5$ mm; $T^* = 403 \pm 1$ K. Solid red line represents the linear fitting: $\hat{r}(\hat{T}) = \gamma + (1-\gamma)\hat{T}$; $\gamma = 2.89$

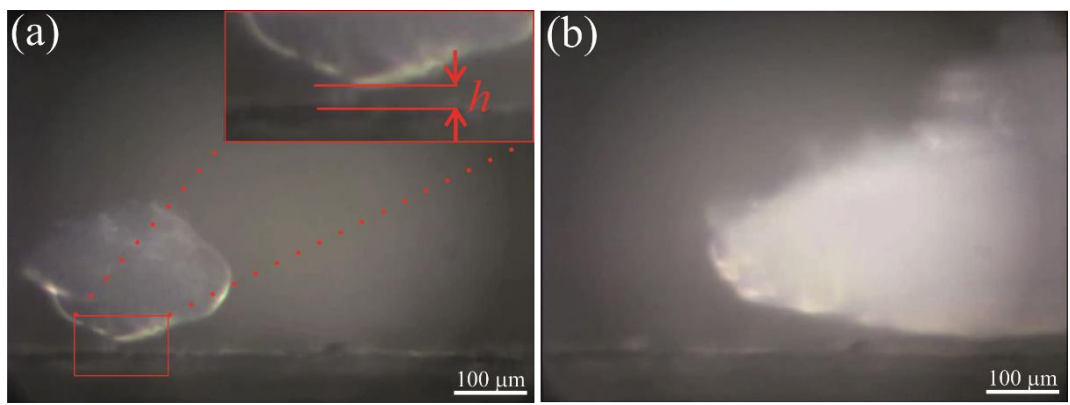

**Fig. 3 | Levitating fumed silica clusters.** Insets (**a**) and (**b**) show levitating fumed silica clusters of different size. The levitation height is established as $h = 18 \pm 1$ µm.

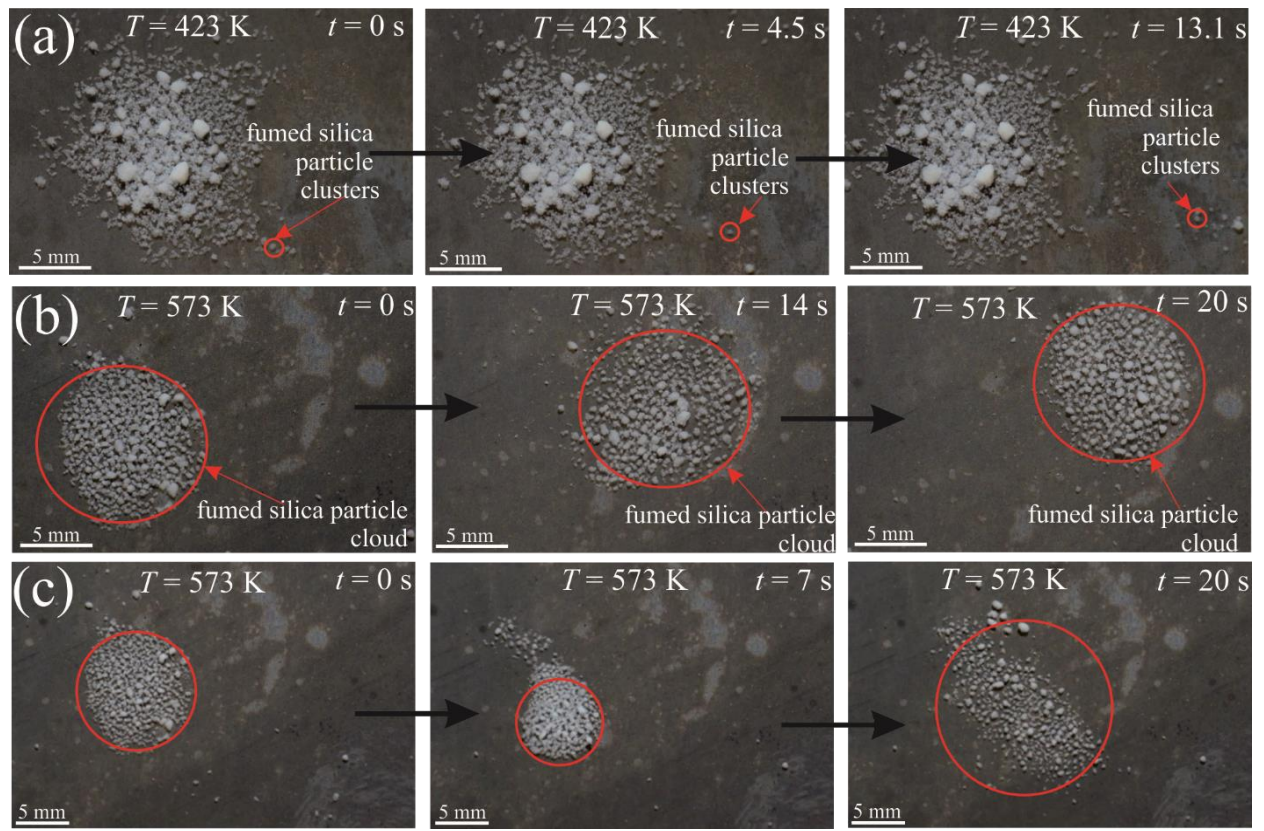

**Fig. 4 |** Structure of the cloud of fluorinated fumed silica particles levitating in the open space is illustrated at different temperatures.

## Methods

### Materials

Powders used in our experiments were the Poly(tetrafluoroethylene) (PTFE) particles (average diameter ~1 μm) supplied by Sigma-Aldrich; lycopodium particles supplied by Fluka; carbon black (Vulcan XC72R) was supplied by Cabot; hydrophobized silica particles (particle size range 20-600 nm) prepared as described in ref. 35. Fumed fluorosilica particles were prepared as described in ref. 36. Among all of the particles, only fumed fluorosilica particles demonstrate levitation at atmospheric pressure. The primary diameter of these particles was 20-30 nm, and they originated from hydrophilic silica after reaction with tridecafluoro-1,1,1,2-tetrahydrooctyltrimethoxysilane. The residual silanol content on their surfaces is 50% and the fluorine content is 10.9%.

### Spreading and Levitation Experiments

Spreading experiments were performed in the closed space as shown in Extended Data Fig. 1 at atmospheric pressure. A Petri dish or desiccator were used as restricting chambers. The diameter of the Petri dish is 7 cm and the height was 1.3 cm. The diameter of the desiccator was 16 cm and the height was 14 cm. The results of the experiments performed with the Petri dish and desiccator were the same.

The levitation experiments were performed on an electric heater in an open lab environment (the humidity was 45-55%) at atmospheric pressure, as shown in Extended Data Fig. 1b. The temperature of the electric heater fluctuated with time. A thick copper block with a dimension of 11.7cm ×9.7cm ×1cm was placed on top of the electric heater surface, thus stabilizing the surface temperature as illustrated with Extended Data Fig. 2. The copper block surface temperature and the surface temperature of this electric heater surface were monitored using a TM-902C Lutron Digital Thermometer with a chromel-alumel thermocouple with an accuracy of ± 1°C and in parallel with the IR thermometer (Extech Instruments, USA) with an accuracy of ± 1°C. The experimental setup is schematically shown in Extended Data Fig. 1b.

Levitation and movement of the particles were captured using a Nikon 1 v3 camera from the top, and the levitation height was measured from the side using a Lieder MZB 930 microscope with DEM200 digital eyepiece.

**Calculation of the Voronoi entropy**

We used MATLAB software to build and process the Voronoi tessellations for levitating clusters and to calculate the corresponding Voronoi entropy values.[33,34] To construct the Voronoi diagrams from the images of the levitating clusters, we used the open access software designed at the Department of Physics and Astronomy at the University of California, Irvine, which is available *via* the website https://www.physics.uci.edu/~foams/do_all.html.

The Voronoi entropy of the given set of points located in a plane is defined as:

$$S_{vor} = -\sum_i P_i \ln P_i, \qquad (17)$$

where $P_i$ is the fraction of polygons possessing *n* edges for a given Voronoi diagram (also called the coordination number of the polygon) and $i$ is the total number of polygon types with different number of edges. The summation in Equation 17 is performed from $i = 3$ to the largest coordination number of any available polygon, *e.g.* to $i = 6$ if a polygon with the largest number of edges is a hexagon.[33,34]

**Calculation of the averaged distance between clusters forming the levitating cloud**

The double time-space averaging procedure was used to calculate the average distance between the clusters constituting the levitating powder clouds. Photo images (frames) of the clusters were treated as follows: at the first stage nearest neighboring clusters were established with the Voronoi diagram technique, as shown in **Extended Data Fig. 7**.

Thus, an average distance between the given *i*th cluster and its nearest neighbors was calculated for the *j*th frame and denoted $\tilde{s}_{ij}(t)$. At the next stage the mean distance between clusters averaged across the given *j*th frame was calculated as:

$$\bar{s}_j(t) = \frac{\sum_{i=1}^{n} s_{ij}(t)}{n}, \tag{18}$$

where *n* is the number of clusters. On the next stage the time averaging was carried across the *m* frames and the double time-space mean distance between clusters was defined as follows:

$$\bar{\bar{s}} = \frac{\sum_{j=1}^{m}\sum_{i=1}^{n} s_{ij}(t)}{mn} = \frac{\sum_{j=1}^{m} \bar{s}_j(t)}{m}. \tag{19}$$

**Data availability:** The data that support the findings of this study are available upon reasonable request from the corresponding author.

**Code availability:** All codes written for use in this study are available upon reasonable request from the corresponding author.

**Acknowledgements:** Edward Bormashenko and Leonid Dombrovsky are grateful to the Russian Science Foundation (project 19-19-00076) for the financial support of the work.

**Author contributions:** Effect revealed, P.K.R.; Conceptualization, P.K.R., S.S., L.A.D and E.B.; P.K.R., I.L. and N.S. designed and the performed experiments; fabrication of hydrophobic silica particles, B.P.B. and V.V.; data curation, P.K.R., I.L.; data analysis, P.K.R., I.L. and N.S.; physical modeling, L.A.D., V.L., A.K. and E.B.; E.B. wrote the manuscript with input from all authors; writing-review and editing done L.A.D., S.S., B.P.B., P.K.R., I.L., N.S. and E.B.; visualization, P.K.R. and E.B.; supervision, S.S. and E.B.; project administration, S.S. All authors have read and agreed to the published version of the manuscript.

**Competing interests:** The authors declare no conflict of interest.

**Correspondence and requests for materials** should be addressed to E.B.

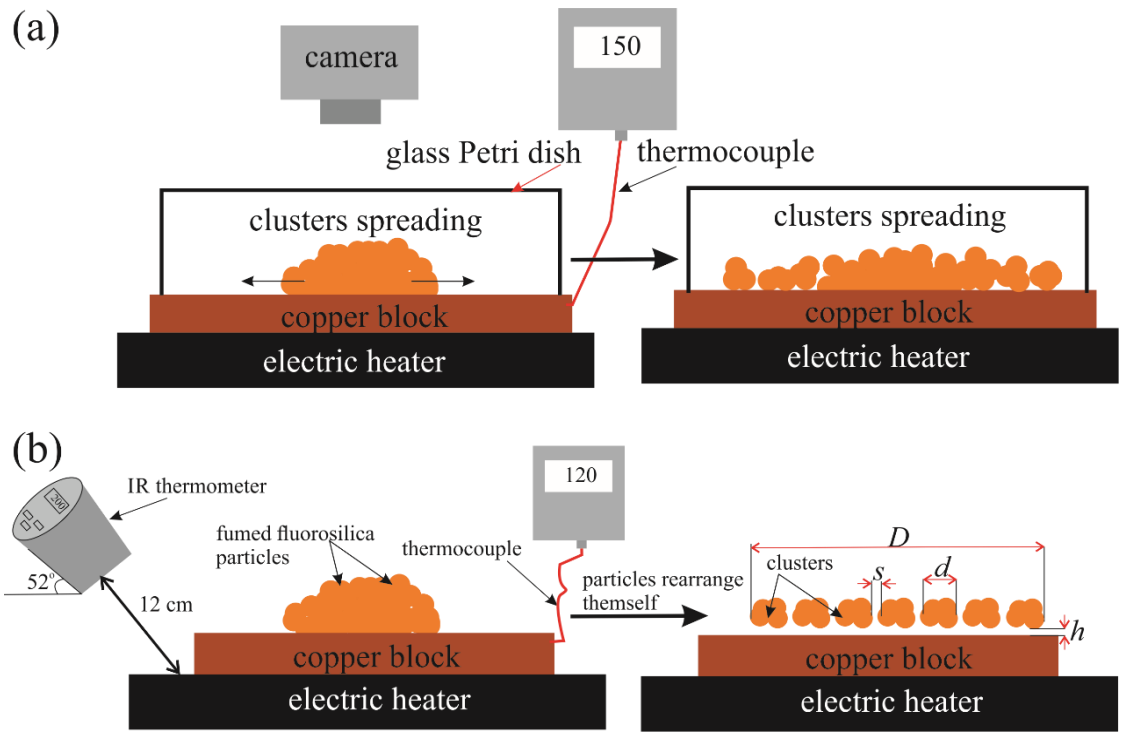

**Extended Data Fig. 1 | Schemes of the experiment performed in the (a) closed and (b) open space are shown.** The geometrical parameters of the cloud and clusters are shown in inset (b).

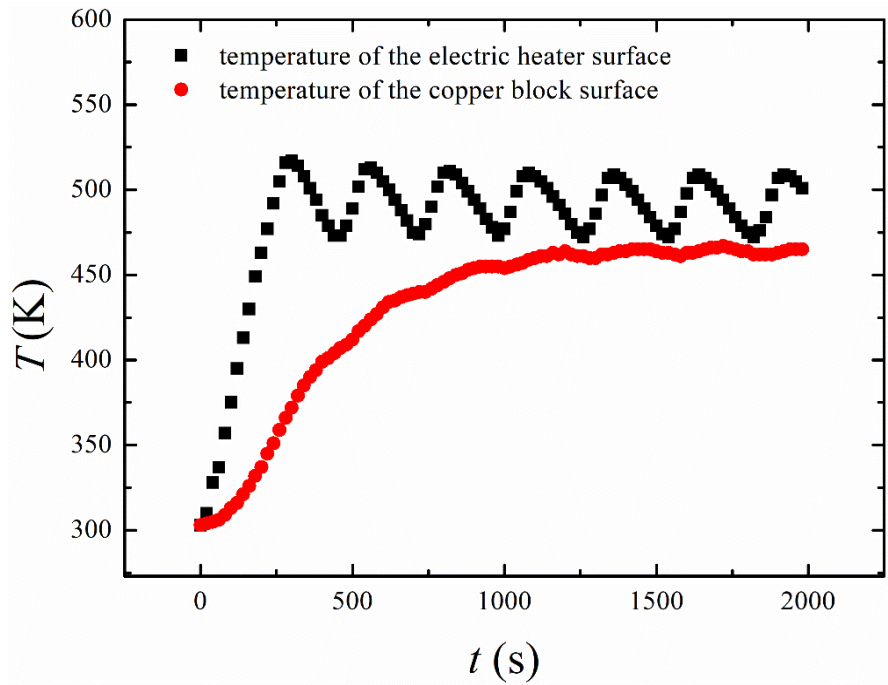

**Extended Data Fig. 2 | Temperature profile of the electric heater surface and the copper block surface.** In the graph black data points represent the electric heater surface temperature measured by IR thermometer and red data points representing the copper block surface temperature as measured by the thermocouple.

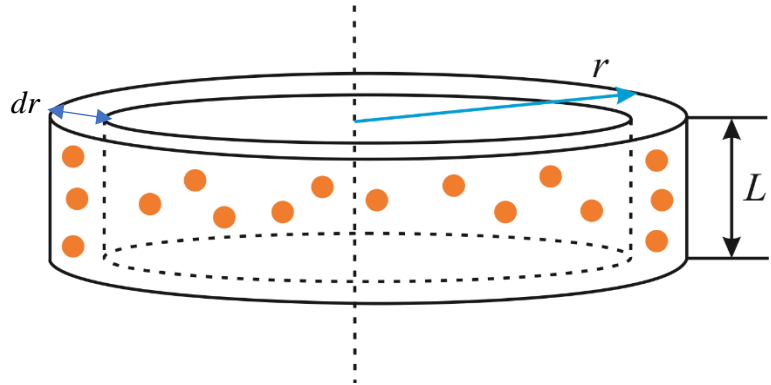

**Extended Data Fig. 3. Cylindrical cloud built of levitating clusters.** Levitating clusters shown with orange spots. Geometrical parameters of the cloud (radius *r* and thickness *L*) are depicted. Compression of the cloud by $dr$ is shown.

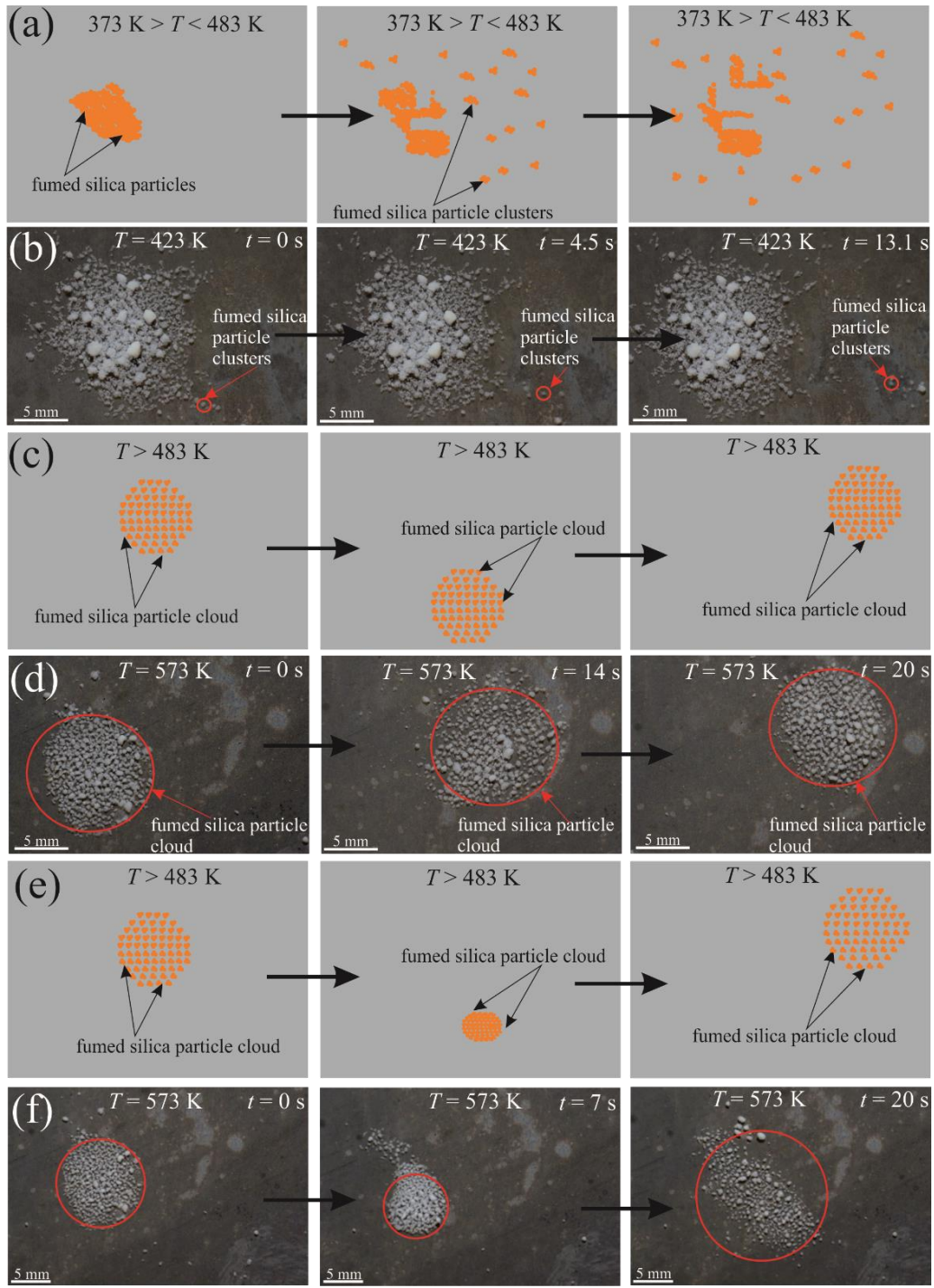

**Extended Data Fig. 4 | The structure of the cloud levitating in the open space is illustrated.**

**a,b,** Clusters constituting the cloud are shown. **c, f,** The cloud and its motion are shown.

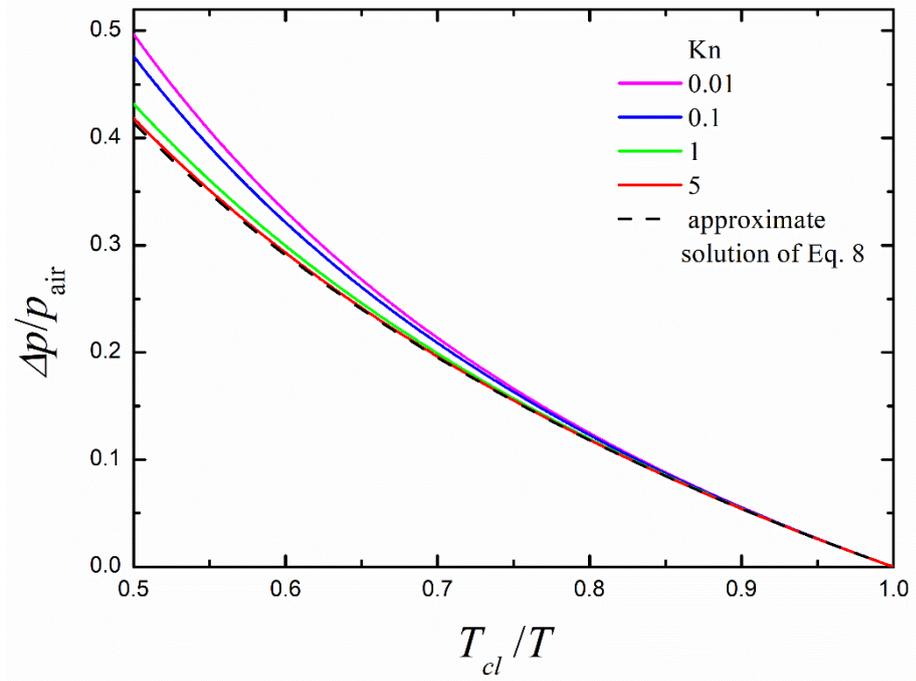

**Extended Data Fig. 5 |** Pressure difference between the lower and upper surfaces of the levitating cloud is shown.

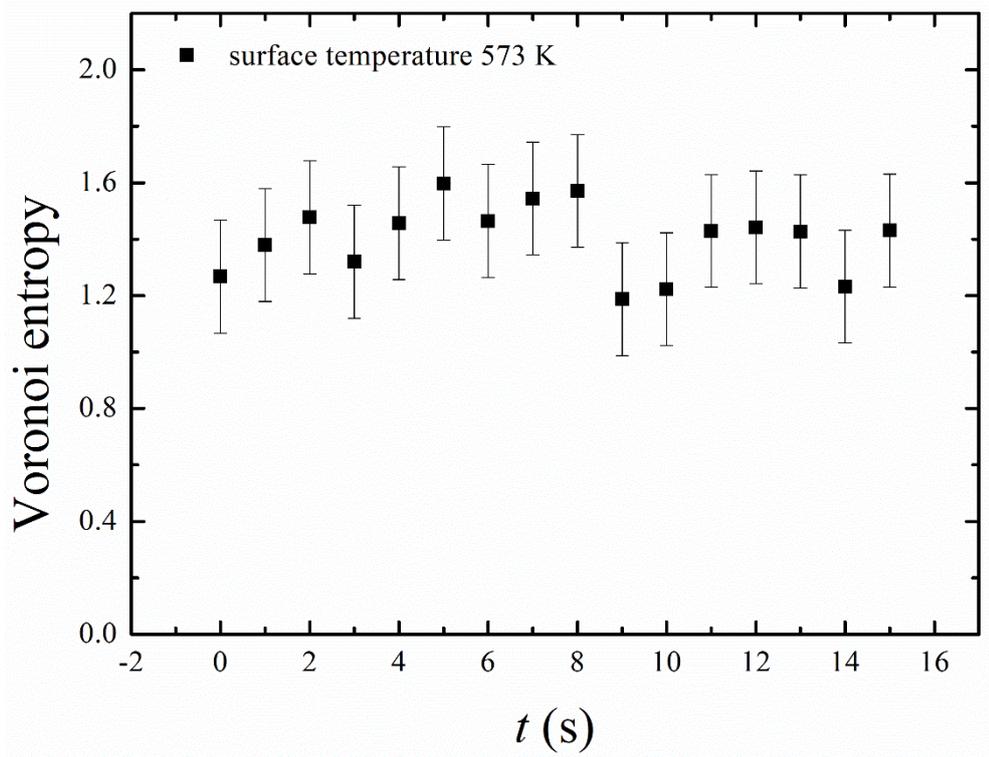

**Extended Data Fig. 6** | The time evolution of the Voronoi entropy of the cloud at $T = 573$ K is depicted.

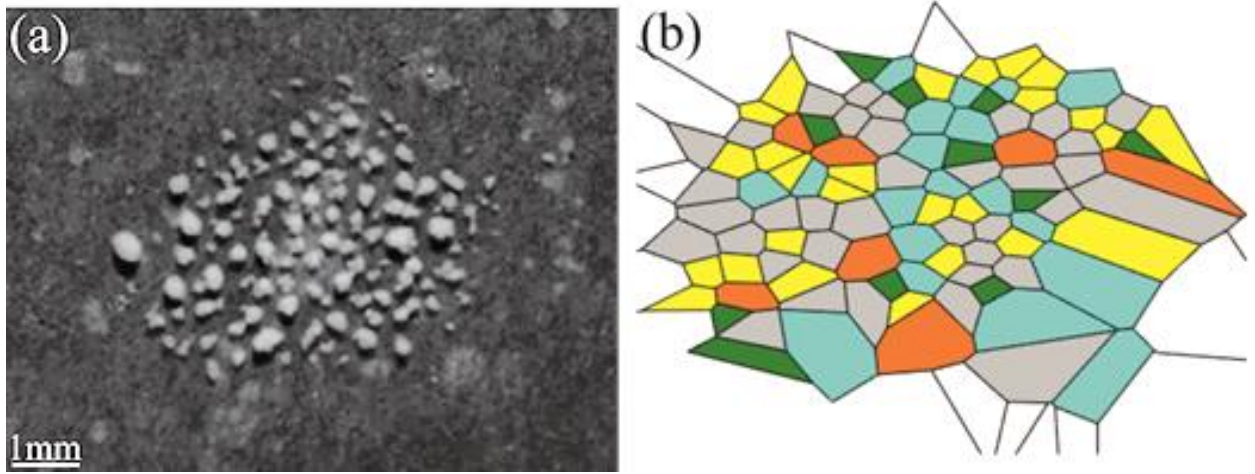

**Extended Data Fig. 7 | Levitating cloud of clusters.** (a) Cluster photo at $T = 573$ K. (b) Voronoi diagram of the cluster. The Voronoi entropy value related to the cluster is $S_{vor} = 1.47$. Color mapping: green polygons - tetragons, yellow – pentagons, grey – hexagons, blue – heptagons, orange – octagons.